\documentstyle[preprint,aps]{revtex}
\begin{document}
\draft
\title{Excitons in two coupled conjugated polymer chains}
\author{Z. G. Yu$^1$, M. W. Wu$^3$, X. S. Rao$^2$, X. Sun$^2$ and A. R. Bishop$^1$}
\maketitle

\vspace{1.0cm}

{\it
\begin{tabular}{ll}
1. & Theoretical Division, Los Alamos National Laboratory,
Los Alamos, NM 87545, USA$^*$\\
2. & T.D. Lee Laboratory of Physics Department, Fudan University,
Shanghai 200433$^*$\\
   & and National Laboratory of Infrared Physics, Shanghai 200083,
   China\\
3. &Department of Physics and Engineering Physics, Stevens
Institute of Technology,\\
  &Hoboken, New Jersey 07030, USA
\end{tabular}
}
\vspace{2.0cm}

Short title: Excitons in two polymer chains\\

\vspace{2.0cm}
Classification Codes: 7135, 7145G
\newpage
\begin{abstract}
We have studied the exciton states in two coupled conjugated polymer
chains which are modeled individually by the Su-Schrieffer-Heeger
Hamiltonian and coupled by an interchain electron-transfer term.
Both the intra- and inter-chain long range Coulomb interactions are
taken into account. The properties of the lowest symmetric and
antisymmetric exciton states are extensively discussed for both the
parallel and anti-parallel ordering between these two chains. It is
found that, for these two kinds of ordering, the features of excitons 
are
quite different.
Possible implications for the experiment of luminescent polymers are
also addressed.
\end{abstract}
\newpage
\section{Introduction}

The significance of excitons in conjugated polymers has been
recognized for many years in a class known as
polydiacetylene \cite{CZ87}. Recently, the
interest in excitons has been heightened by discovering that
poly({\em p-}phenylene vinylene) (PPV) and its derivatives can be
used as the
active luminescent layer in electroluminescent light-emitting diode
devices \cite{BBBM},
since it is believed that radiative recombination of singlet
excitons gives rise to luminescence. Although a general picture
of understanding of photoinduced absorption (PA) and
photoconductivity
(PC) experiments of PPV and its derivatives remains a subject of
intense debate \cite{FLSK,LYMH,LJWB,YRPG,HYJR,CMJW,CC93,me,GCM},
the dramatically
different PA behavior of dilute solutions and thin films of
poly[2-methoxy, 5-(2' ethyl-hexoxy)-1,4 phenylene vinylene]
(MEH-PPV) \cite{Yan},
together with the photoconductivity being observed to start at the
absorption edge in PPV \cite{PHBB}, clearly
indicate the important role the interchain coupling plays.
Actually, electron-diffraction experiments indicate that the
interchain
coherence length of PPV is about 60 \AA \cite{HBBF}.
However, the properties of excitons in coupled chains, to the best
of our knowledge, have never been discussed.

There have been several works to develop the exciton theory in a
single polymer chain \cite{Su84,HN85,TIH87,AYS92}.
Abe and co-workers introduced the standard exciton theory \cite{NTA}
to one-dimensional polymers \cite{AYS92}.
Within the single configuration interaction
(SCI) approximation, the energy levels and wave functions of exciton
states in a long chain can be obtained. In this paper, we will extend
their approach and explore the features of excitons in two coupled
polymer chains.
Interestingly enough, we find that, for two kinds of ordering, i.e.,
the parallel-
and anti-parallel- ordering between the two chains, the properties of
excitons are quite different. Although this study does not result in
quantitative explanations of the PA and PC experiments,
it is helpful for understanding the
photophysics better in luminescent polymers like PPV, and is also a
foundation
of further calculations of optical properties of coupled polymers. In
Sec. II, the model is defined and the formulation is derived. In Sec.
III, we present numerical results on the lowest symmetric and
antisymmetric excitons and discuss some implications for experiments.

\section{Formalism}

We start with the two coupled chains model introduced by Baeriswyl and
Maki \cite{BM83},
in which, each chain is described by the Su-Schrieffer-Heeger
Hamiltonian (SSH) \cite{HKSS}
\begin{equation}
H_j=-t\sum_n[1-(-1)^n z_j](c^{\dag}_{jn+1}c_{jn}+{\rm H.c.})~,
\end{equation}
where $j=1,2$ denotes the chain index, and is coupled by an interchain
hopping term
\begin{equation}
H_{\perp}=-t_{\perp}\sum_n(c^{\dag}_{1n}c_{2n}+{\rm H.c.})~.
\end{equation}
Here $c_{jn}$ is the annihilation operator of the electron at site $n$
on the $j$th chain, the spin indices have been omitted for simplicity.
Each chain is assumed to be dimerized in accordance with the
Peierls theorem \cite{Pe55},
$z_j$ is the dimerization amplitude of the $j$th chain,
$|z_1|=|z_2|=z$, but they may differ in sign.
Strictly speaking, the SSH model is directly applicable only to 
polyacetylene,
however, recent works have shown that the primary excitation in
luminescent polymers like PPV can also be described within the linear
chain model \cite{SEGR,me,GCM}. In PPV and its derivatives, the lowest
excitonic wave function extends over several repeat units
\cite{CC93,PHBB},
the properties of exciton are therefore not very sensitive to the 
delicate
structure
within the unit. From the viewpoint of renormalization \cite{FGP96}, 
we can
map the complex structure of PPV into an effective SSH system with the
same significant physical properties by integrating out the freedom of
benzene rings and only considering the electrons on the nonbenzene 
carbon
atoms \cite{me}. Thus the
features of exciton in SSH model must have some implication to that in
luminescent polymers. We have also adopted the rigid lattice
approximation,
since many experiments and theories had demonstrated that the primary
excitation is the exciton and electron-electron interactions
are predominant over electron-lattice interactions in luminescent
polymers \cite{LJWB,HYJR,CMJW,me,GCM,SEGR}. Theoretical works that 
have
taken into account the electron-lattice interaction also show that
incorporation of lattice relaxation effect would not lead to an 
increase
in binding energy of exciton \cite{shuai}.
This simplification enables us to handle electron-electron
interactions in long
chains and arrive at an understanding of {\em electronic states} in
luminescent polymers without loss of essential physics, although the
quantitative explanation of some {\em lattice} property like vibronic
structure or bond length should, indeed, take into account
the lattice relaxation effects \cite{BSFB}.

By introducing operators $a_{jk}$ and $b_{jk}$ through the
relation
(we take the lattice constant $a=1$ in this section) \cite{BM83}
\begin{equation}
c_{jn}=\frac{1}{\sqrt{N}}\sum_k e^{ikn}[(-1)^n a_{jk}+i b_{jk}]~,
\label{cab}
\end{equation}
here $N$ is the number of sites per chain, and by using the Bogoliubov
transformation
\begin{equation}
\left( \begin{array}{cc}
a_{jk} \\
b_{jk} \\
\end{array}  \right)
=\left( \begin{array}{cc}
\cos \theta_{jk}   &   \sin \theta_{jk} \\
-\sin \theta_{jk}   &   \cos \theta_{jk} \\
\end{array}  \right)
\left( \begin{array}{cc}
\alpha_{jk}  \\
\beta_{jk} \\
\end{array}  \right)~~,
\label{bog}
\end{equation}
$H_j$ is diagonalized when $\tan 2\theta_{jk}=z_j \tan k$,
\begin{equation}
H_j=\sum_k
E_k(\alpha^{\dag}_{jk}\alpha_{jk}-\beta^{\dag}_{jk}\beta_{jk})~,
\end{equation}
with
\begin{equation}
E_k=2t\sqrt{\cos^2 k+z^2_j \sin^2 k}~.
\end{equation}
The interchain hopping term then becomes
\begin{equation}
H_{\perp}=-t_{\perp}\sum_k[\cos(\theta_{1k}-\theta_{2k})
(\alpha^{\dag}_{1k}\alpha_{2k}+\beta^{\dag}_{1k}\beta_{2k})
+\sin (\theta_{1k}-\theta_{2k})(\beta^{\dag}_{1k}\alpha_{2k}-
\alpha^{\dag}_{1k}\beta_{2k})+{\rm H.c.}]~.
\end{equation}

There are two kinds of orderings between the two chains, the parallel
ordering ($\theta_{1k}=\theta_{2k}$) and anti-parallel one
($\theta_{1k}=-\theta_{2k}$) \cite{BM83}.
For the case of $\theta_{1k}=\theta_{2k}$,
\begin{equation}
H_{\perp}=-t_{\perp}\sum_k(\alpha^{\dag}_{1k} \alpha_{2k}
+\beta^{\dag}_{1k}\beta_{2k}+{\rm H.c.})~,
\end{equation}
the full Hamiltonian $H=H_1+H_2+H_{\perp}$ can be written as
\begin{equation}
H=\sum_k (\alpha^{\dag}_{1k}~~\alpha^{\dag}_{2k})
\left( \begin{array}{cc}
E_{k}  &  -t_{\perp}  \\
-t_{\perp} &  E_k \\
\end{array}  \right)
\left( \begin{array}{cc}
\alpha_{1k}  \\
\alpha_{2k} \\
\end{array}  \right)+
\sum_k (\beta^{\dag}_{1k}~~\beta^{\dag}_{2k})
\left( \begin{array}{cc}
-E_{k}  &  -t_{\perp}  \\
-t_{\perp} &  -E_k \\
\end{array}  \right)
\left( \begin{array}{cc}
\beta_{1k}  \\
\beta_{2k} \\
\end{array}  \right)~,
\label{hpa}
\end{equation}
and is readily diagonalized by the orthogonal transformation
\begin{equation}
(A_{1k}~A_{2k}~B_{1k}~B_{2k})^T=\b{O}(\alpha_{1k}~\alpha_{2k}~
\beta_{1k}~\beta_{2k})^T~,
\label{oth}
\end{equation}
\begin{equation}
H=\sum_k[(E_k-t_{\perp})(A^{\dag}_{1k}A_{1k}-B^{\dag}_{1k}B_{1k})
+(E_k+t_{\perp})(A^{\dag}_{2k}A_{2k}-B^{\dag}_{2k}B_{2k})]~,
\end{equation}
where, $A^{\dag}_{ik}$ and $B^{\dag}_{ik}$ ($i=1,2$) create an
electron in the $i$th conduction and valence band respectively.
  
For the case of $\theta_{1k}=-\theta_{2k}=\theta_k$,
\begin{equation}
H_{\perp}=-t_{\perp}\sum_k[\cos 2\theta_k(\alpha^{\dag}_{1k}
\alpha_{2k}
+\beta^{\dag}_{1k}\beta_{2k})+\sin
2\theta_k(\beta^{\dag}_{1k}\alpha_{2k}-\alpha^{\dag}_{1k}\beta_{2k})
+{\rm H.c.}]~,
\end{equation}
the total Hamiltonian reads
\begin{equation}
H=\sum_k
(\alpha^{\dag}_{1k}~\alpha^{\dag}_{2k}~\beta^{\dag}_{1k}
~\beta^{\dag}_{2k})
\left( \begin{array}{cccc}
E_{k}  &  -t_{\perp}\cos2\theta_k & 0  & t_{\perp}\sin2\theta_k \\
-t_{\perp}\cos2\theta_k & E_k &  -t_{\perp}\sin2\theta_k & 0 \\
0 & -t_{\perp}\sin2\theta_k & -E_k & -t_{\perp}\cos2\theta_k \\
t_{\perp}\sin2\theta_k & 0 & -t_{\perp}\cos2\theta_k & -E_k \\
\end{array}  \right)
\left( \begin{array}{cc}
\alpha_{1k}  \\
\alpha_{2k} \\
\beta_{1k}\\
\beta_{2k}\\
\end{array}  \right)~.
\label{hanp}
\end{equation}

Making the orthogonal transformation $\b{O}$, we
obtain the diagonalized Hamiltonian,
\begin{equation}
H=\sum_k[\varepsilon_{1k}(A^{\dag}_{1k}A_{1k}-B^{\dag}_{1k}B_{1k})
+\varepsilon_{2k}(A^{\dag}_{2k}A_{2k}-B^{\dag}_{2k}B_{2k})]~,
\end{equation}
with
\begin{equation}
\varepsilon_{1k}=\sqrt{E^2_k+t^2_{\perp}+4t t_{\perp} \cos k}~,
\end{equation}
\begin{equation}
\varepsilon_{2k}=\sqrt{E^2_k+t^2_{\perp}-4t t_{\perp} \cos k}~,
\end{equation}
and, meanwhile, the transformation matrix $\b{O}$.
 
We add the long range Coulomb interaction $H_{e-e}$ to $H$ and study
the exciton state,
\begin{equation}
H_{e-e}=\frac{1}{2}\sum_{IJll'}\sum_{ss'}V^{IJ}_{ll'}\rho_{Ils}
\rho_{Jl's'}~,
\end{equation}
where $\rho_{Ils}=c^{\dag}_{Ils}c_{Ils}-1/2$, $I$ and $J$ are the
chain indices, and the interaction potential
$$
V^{IJ}_{ll'}=\left \{
\begin{array}{ll}
V_{ll'}~,        &  I=J \\
\tilde{V}_{ll'}~,  &  I\ne J~.
\end{array}
\right.
$$
The intrachain Coulomb interaction is the commonly used
form \cite{AYS92}
$$
V_{ll}=U,~~V_{ll'}=\frac{V}{|l-l'|}(l\ne l')~.
$$
Here $U$ is the on-site Hubbard repulsion, $V$ the nearest-
neighbor Coulomb interaction. We assume that the interaction potential
between the two chains has a relatively simple form,
$$
\tilde{V}_{ll'}=\frac{\tilde{U}}{\sqrt{R^2_0+(l-l')^2}}~,
$$
$R_0$ will be set to be 1 in following calculations.

The procedure to determine the exciton state is a standard one.
First, we use the single particle
state of $H$ to construct the ground state $|g \rangle$, $|g \rangle=
\prod_{jk}B^{\dag}_{jk\uparrow}B^{\dag}_{jk\downarrow}|0 \rangle$.
Within the SCI
approximation, the exciton state can be achieved by diagonalizing the
Hamiltonian $H+H_{e-e}$ in the subspace of the single electron-hole
excitation
\begin{equation}
|i,k_c;j,k_v
\rangle \equiv\frac{1}{\sqrt{2}}(A^{\dag}_{ik_c\uparrow}
B_{jk_v\uparrow}
\pm A^{\dag}_{ik_c\downarrow}B_{jk_v\downarrow})|g \rangle~,
\end{equation}
where $+$ is for the spin singlet, $-$ for one of the triplet. $k_c$
and $-k_v$ are momenta of the electron in conduction band
and the hole in valence band, respectively. $i$ and $j$ $(=1,2)$
are the band indices.

For the spin singlet exciton,
\begin{eqnarray}
&&\langle i',k'_c;j',k'_v|H+H_{e-e}-E_0|i,k_c;j,k_v \rangle\nonumber\\
&=&\delta_{k'_v,k_v}\delta_{k'_c,k_c}[\delta_{jj'}
(\delta_{ii'}\varepsilon_{ik_c}
+\sum_{IJll'}V^{IJ}_{ll'} \langle A_{i'k_c} c^{\dag}_{Jl'} \rangle
\langle c_{Jl'} c^{\dag}_{Il} \rangle
\langle c_{Il} A^{\dag}_{ik_c} \rangle)\nonumber\\
&+&\delta_{ii'} (-\delta_{jj'}\varepsilon_{jk_v}
-\sum_{IJll'} V^{IJ}_{ll'}
\langle B^{\dag}_{j'k_v} c_{Jl'} \rangle
\langle c_{Il} c^{\dag}_{Jl'} \rangle
\langle c^{\dag}_{Il} B_{jk_v} \rangle)]+2E_X-E_C~,
\label{sing}
\end{eqnarray}
where $E_0=\langle g|H+H_{e-e}|g \rangle$, and for the triplet,
\begin{eqnarray}
&&\langle i',k'_c;j',k'_v|H+H_{e-e}-E_0|i,k_c;j,k_v \rangle\nonumber\\
&=&\delta_{k'_v,k_v}\delta_{k'_c,k_c}[\delta_{jj'}
(\delta_{ii'}\varepsilon_{ik_c}
+\sum_{IJll'} V^{IJ}_{ll'} \langle A_{i'k_c} c^{\dag}_{Jl'} \rangle
\langle c_{Jl'} c^{\dag}_{Il} \rangle
\langle c_{Il} A^{\dag}_{ik_c} \rangle)\nonumber\\
&+&\delta_{ii'} (-\delta_{jj'}\varepsilon_{jk_v}
-\sum_{IJll'} V^{IJ}_{ll'}
\langle B^{\dag}_{j'k_v} c_{Jl'} \rangle
\langle c_{Il} c^{\dag}_{Jl'} \rangle
\langle c^{\dag}_{Il} B_{jk_v} \rangle)]-E_C~,
\label{trip}
\end{eqnarray}
and
\begin{equation}
E_X=\sum_{IJll'}V^{IJ}_{ll'}\langle A_{i'k'_c} c^{\dag}_{Il} \rangle
\langle c_{Jl'} A^{\dag}_{ik_c} \rangle
\langle B^{\dag}_{j'k'_v} c_{Il} \rangle
\langle c^{\dag}_{Jl'} B_{jk_v} \rangle~,
\end{equation}
\begin{equation}
E_C=\sum_{IJll'}V^{IJ}_{ll'}\langle A_{i'k'_c} c^{\dag}_{Jl'} \rangle
\langle c_{Jl'} A^{\dag}_{ik_c} \rangle
\langle B^{\dag}_{j'k'_v} c_{Il} \rangle
\langle c^{\dag}_{Il} B_{jk_v} \rangle~,
\end{equation}
where $\langle ... \rangle \equiv \langle g|..|g \rangle$. We show how
to evaluate $\langle c^{\dag}_{Il} A_{ik_c} \rangle$,
$\langle B^{\dag}_{jk_v} c_{Il} \rangle$, and
$\langle c^{\dag}_{Jl'} c_{Il} \rangle$ in the Appendix.

This two-coupled-chain system also has the symmetry with respect to
the spatial inversion at a bond center like a single chain.
The inversion operator $R$ is defined by
\begin{equation}
R[c_{jn}]=c_{jN-n+1}~,
\end{equation}
and it is easy to prove
$$
R[\alpha^{\dag}_{jk}]=-e^{ik}\alpha^{\dag}_{j-k}~,~
R[\beta_{jk}]=e^{-ik}\beta_{j-k}~,
$$
so
\begin{equation}
\alpha^{\dag}_{ik_c}\beta_{jk_v}|g \rangle \stackrel{R}{\to}
-e^{i(k_c-k_v)}\alpha^{\dag}_{-ik_c}\beta_{j-k_v}|g \rangle~,
\end{equation}
and from the transformation matrix $\b{O}$, which diagonalizes the
Hamiltonians (\ref{hpa}) and (\ref{hanp}), we also have
\begin{equation}
|i,k_c;j,k_v \rangle \stackrel{R}{\to}
-e^{i(k_c-k_v)}|i,-k_c;j,-k_v \rangle~.
\end{equation}

Thus we can construct the symmetric state ($A$) and anti-symmetric
one ($B$). The $A$ state is written as
\begin{equation}
|i,k_c;j,k_v;- \rangle
=\frac{1}{\sqrt{2}}(|i,k_c;j,k_v \rangle -
e^{i(k_c-k_v)}|i,-k_c;j,-k_v \rangle)~,
\end{equation}
and the $B$ state is
\begin{equation}
|i,k_c;j,k_v;+ \rangle
=\frac{1}{\sqrt{2}}(|i,k_c;j,k_v \rangle +
e^{i(k_c-k_v)}|i,-k_c;j,-k_v \rangle)~.
\end{equation}
In the numerical calculation, we may confine ourselves in the $A$ or
$B$ subspaces to
diagonalize the Hamiltonian $H+H_{e-e}$ since the matrix element
between $A$ and $B$ states vanish.
In the exciton state, the relative and center-of-mass motion can also
be separated by introducing the variable $k$ and $K$ so that $k_c=k+K$
and $k_v=k-K$. $k$ and $2K$ are the momenta of the relative motion and
center-of-mass motion of the electron-hole pair. Since we are only
interested in the relative motion, we will just consider the
case of $K=0$, and now the basis is $|i,k;j,k;\pm \rangle$.

The wave function of exciton in real space can be determined by
\begin{equation}
\psi(I,n;J,l;\pm)
=\sum_{ij,k}
\langle I,n,\uparrow;J,l,\downarrow|i,k;j,k;\pm \rangle
\langle i,k;j,k;\pm|\psi \rangle~,
\end{equation}
the positions of electron and hole are at site $n$ on the $I$th chain
and at site $l$ on the $J$th chain, respectively.
$\langle i,k;j,k;\pm|\psi \rangle$ can be obtained by diagonalizing
the matrices (\ref{sing}) and (\ref{trip}), and
\begin{eqnarray}
&&\langle I,n,\uparrow;J,l,\downarrow|i,k;j,k;\pm \rangle\nonumber\\
&=&\frac{1}{\sqrt{2}}(\langle c_{In}A^{\dag}_{ik} \rangle
\langle c^{\dag}_{Jl}B_{jk} \rangle \pm
\langle c_{In}A^{\dag}_{i-k} \rangle
\langle c^{\dag}_{Jl}B_{j-k} \rangle )~.
\end{eqnarray}

\section{Numerical results and discussions}

We have carried out numerical calculations to show the properties of
excitons in coupled chains. The intrachain parameters are fixed with
realistic values, $U=2t$, $V=t$, and $z=0.2$.
The interchain coupling $t_{\perp}$ can be
estimated according to the results of the LDF calculations for PPV by
Vogl and Campbell \cite{VC90}. In PPV, the total interchain coupling
for a pair of monomers $t_{\rm mon}=0.64$ eV \cite{MC95}.
Naturally, we set the total coupling
for a pair of unit in the SSH model $2t_{\perp}$ to be $t_{\rm mon}$.
Thus $t_{\perp}$ for PPV ranges from $0.12t$ to $0.15t$. In PPV, the 
nearest
interatom distance on adjacent chains is 2.495 \AA \cite{MC95}, while 
the
bond lengths are around 1.4 \AA, so the Coulomb interaction strength
$\tilde{U}=0.5t$ is a reasonable estimate for PPV.
We will also vary the interchain
hopping $t_{\perp}$ and interchain Coulomb interaction parameter
$\tilde{U}$ to make the interchain effects more transparent.
The system we study consists of two chains of $N=200$. The exciton
wave functions in the following figures represent the relative
distribution of the hole in different positions when
the electron is located at site $n=101$ on the first chain. The
$200+i$th site in figures means the $i$th site of the second chain.

First we study the case of parallel ordering. 
Figure 1 illustrates the wave functions of the lowest $^1 B$ and $^1 
A$
states with $t_{\perp}=0.15t$ and $\tilde{U}=0.5t$. Since the actual
wave function contains rapid staggered oscillation with a period of
$4a$, we have plotted $\psi(I=1,n=101;J,l;\pm)/f(l)$ with
$f(l)=\sqrt{2}\cos[\frac{\pi}{2}(l+\frac{1}{2})]$ for the odd number
$l$ \cite{AYS92}. We give in Tables I and II the exciton energy
$E_{\rm ex}$
and the probability $P$ that the hole is at the second chain for
several groups of $t_{\perp}$ and $\tilde{U}$.
>From these tables, we can see that the hole has more probability to
stay at the second chain for both $B$ and $A$ states as
the interchain hopping $t_{\perp}$ and interchain Coulomb
interaction $\tilde{U}$ increase.
Since the energies of the $A$ states
are close to the continuum band and the wave functions nearly extend
over the
whole system, the amplitude of wave function in the second chain is
comparable to that of the first chain. The energies of $B$ states
sit, however, deep in the gap, so that the amplitude of wave function
in the second chain is smaller and more sensitively depends on the
interchain couplings than
the case of $A$ states. The triplet has similar features of the
singlet.

Now we divert our attention to the case of anti-parallel ordering,
which is more likely to be realized in practical polymers \cite{FCHM}.
Meanwhile
the features of the exciton in this case are more interesting.
We describe
the lowest $^1 B$ and $^1 A$ states for
$t_{\perp}=0.15t$ and $\tilde{U}=0.5t$ in Fig. 2. We have plotted
$\psi(I=1,n=101;J,l;\pm)/f(l)$ for the odd number $l$ when $J=1$ but
for the even number $l$ when $J=2$. The exciton
energy $E_{\rm ex}$, the probability $P$ that the hole stays in the
second chain, and the separation $d$ between positions of maxima
of wave function amplitude in two chains are listed in Tables III and
IV for several groups of the interchain parameters.
It is shown that the larger $\tilde{U}$ leads to the more chance
to find the hole in the second chain. However,
the dependence of probability $P$ on the
interchain coupling $t_{\perp}$ is not so simple as the case of 
parallel
ordering. For the $^1 B$ state, $P$ becomes larger when $t_{\perp}$
increases, while for the $^1 A$ state, in contradiction with 
intuition,
$P$ decreases as $t_{\perp}$ increases. Unlike the parallel
ordering, where the profiles of wave functions in two chains are
alike and the maxima of wave function amplitude in two chains sit
at the same position, in the case of anti-parallel ordering, the wave
functions in two
chain are well different from each other. For $^1 B$ states, when
the hole approaches the position of electron $n=101$ on the first
chain, the wave function reaches the maximum; but on the second
chain, the wave
function goes to zero when the hole is nearest to the electron. While,
for $^1 A$ states, when the hole and electron overlap on the first
chain, the wave function vanishes; but when the hole is on the second
chain, the wave function reaches the maximum as the hole approaches
the position of electron. The positions with the largest wave function
amplitudes in two chains have a separation $\sim 5-9a$
for both $B$ and
$A$ states. The interchain Coulomb interaction can reduce this
separation to a certain extent.  This big difference between the two
ordering case is understandable.
>From Eqs.\ (\ref{hpa}) and (\ref{hanp}), we can see that,
in the case of
parallel ordering, there is no mixing between valence and
conduction bands of different chains and the system is more
like a two-independent-chain
system; in the case of anti-parallel ordering, however, the valence-
and conduction- band states of these two chains are mixed together and
the two chains are really {\em coupled}.

For PPV, there are two kinds of ordering between adjacent chains,
namely, in-phase and out-of-phase ordering \cite{CWMK,ZBH93,BFHM}.
When the practical
structure of PPV is mapped into the simplified SSH model as stated
before, each type of ordering is neither parallel nor
anti-parallel and the wave function of exciton in PPV should 
therefore be
the composition of that in parallel and anti-parallel ordering. For 
the
$^1 B$ state, from Figs. 1(a) and 2(a), the maximum of the composed
wave function in the first chain should sit at the positon of the
electron, since the profiles of wave function in parallel and
anti-parallel ordering are same; while, in the second chain, the
position of maximum for the composed wave function must be situated
between the site facing the electron and the site where the maximum in
the second chain occurs for anti-parallel ordering.
For the $^1 A$ state, from Figs. 1(b) and 2(b), the composed wave 
function
in the first chain will vanish at the position of the electron, and 
reach
the maximum at the site which lies between the position of the 
electron
and that of the maximum for parallel ordering.
The shape of composed wave function, which shows that the position of
maximum in the second chain deviates from the location of the electron
in the first chain,
implies that if the interchain exciton is produced, the electron and 
hole
tend to be separated with several lattice constant. This is similar 
as the
concept of ``spatially indirect exciton'' proposed by Yan and
co-workers to interpret the PA spectrum in PPV \cite{YRPG,HYJR}.
We emphasize that it is the exchange effect which prevents the 
electron
and hole in different chains from approaching each other in the 
presence
of the interchain electron-electron interaction, and previous 
treatments
to the interchain Coulomb interaction, in which, only the
electrostatic energy is included \cite{MC94}, cannot predict
this feature. Another interesting property is that,
for the $A$ state, the interchain exciton is even more likely to be
created than the intrachain exciton.
Recently, it was documented that 80\% -- 90\% photoexcitations
in PPV are interchain excitations \cite{YRPG},
since in practical materials, defects, interfaces, and thermal 
fluctuation
can lead to the charge transfer and the mixing between the $A$ and 
$B$ states,
the $A$ states with
large possibilities of interchain excitons seem to have
contributions to the great quantity of the interchain excitations
in PPV. Obviously, these interchain excitations are also important to
the PC, which is thought an interchain process in PPV.

In summary, we have studied the excitons in two coupled polymer
chains. The wave functions of the lowest $A$ and $B$ states for
both parallel and anti-parallel ordering have been illustrated. We
have shown the pronounced difference of the property of exciton
between these two kinds of ordering.
For the realistic PPV, whose properties should be the combination of 
that
in each ordering, the electron
and hole tend to be separated with several lattice constant when they
are not in the same chain, and in the $A$ state, the probability
that the electron and hole are in different chains is even larger than
that they are in the same chain. These features are helpful to clarify
the controversies of interpreting experiments of luminescent polymers
like PPV.

\acknowledgments
This work was partly supported by the National Natural Science
Foundation and the Advanced Materials Committee of China. Part of
numerical calculation of the present work was carried out when Z.G.Y.
was visiting Tianjin, China, in the vacation. He thanks Yi Hou for her
hospitality and help during his stay there.
One of the authors (M.W.W) was supported by U.S. Office Naval Research
(Contract No. N66001-95-M-3472) and the U.S. Army Research Office
(Contract No. DAAH04-94-G-0413).

\appendix
\section*{}

In this appendix, we show how to calculate
$\langle c^{\dag}_{Il} A_{ik_c} \rangle$,
$\langle B^{\dag}_{jk_v} c_{Il} \rangle$, and
$\langle c^{\dag}_{Jl'} c_{Il} \rangle$. It is noted that
\begin{equation}
\langle c_{Il}A^{\dag}_{ik_c} \rangle=\langle Il|A^{\dag}_{ik_c}|0
\rangle ~,
\end{equation}
\begin{equation}
\langle B^{\dag}_{jk_v}c_{Il} \rangle=\langle Il|B^{\dag}_{ik_v}|0
\rangle ~,
\end{equation}
and
\begin{equation}
\langle c^{\dag}_{Jl'}c_{Il} \rangle=\sum_{ik_v}
\langle 0|B_{ik_v}|Jl'\rangle \langle Il|B^{\dag}_{ik_v}|0 \rangle~.
\end{equation}

Since from Eqs. (\ref{cab}) and (\ref{bog}), we have
\begin{equation}
\left( \begin{array}{cc}
\langle 2n|\alpha^{\dag}_{jk}|0 \rangle  &
\langle 2n+1|\alpha^{\dag}_{jk}|0 \rangle   \\
\langle 2n|\beta^{\dag}_{jk}|0 \rangle  &
\langle 2n+1|\beta^{\dag}_{jk}|0 \rangle   \\
\end{array}  \right)
=\frac{1}{\sqrt{N}}e^{ik2n}\left( \begin{array}{cc}
\exp(-i\theta_{jk}) & -\exp(i\theta_{jk}+ik)  \\
i\exp(-i\theta_{jk}) & i\exp(i\theta_{jk}+ik)  \\
\end{array}  \right)~,
\end{equation}
with
\begin{equation}
\cos
\theta_{jk}=\frac{1}{\sqrt{2}}(1+\frac{\epsilon_k}{E_k})
^{\frac{1}{2}}~,
\end{equation}
\begin{equation}
\sin
\theta_{jk}=\frac{z_j\sin k}{|z_j\sin k|}
\frac{1}{\sqrt{2}}(1-\frac{\epsilon_k}{E_k})^{\frac{1}{2}}~,
\end{equation}
here $\epsilon_k=2t \cos k$,
then combining Eq. (A4) with the transformation matrix $\b{O}$,
we can obtain $\langle c^{\dag}_{Il} A_{ik_c} \rangle$,
$\langle B^{\dag}_{jk_v} c_{Il} \rangle$, and
$\langle c^{\dag}_{Jl'} c_{Il} \rangle$.

\begin{figure}
\caption{Wave functions of the lowest $^1 B$ and $^1 A$ exciton
states for the interchain hopping $t_{\perp}=0.15t$ and interchain
Coulomb interaction $\tilde{U}=0.5t$ in the case of parallel ordering
between the two chains. (a) is for $^1 B$ state and (b) is for
$^1 A$ one.}
\end{figure}
\begin{figure}
\caption{Wave functions of the lowest $^1 B$ and $^1 A$ exciton
states for $t_{\perp}=0.15t$ and $\tilde{U}=0.5t$ in the case of
anti-parallel
ordering. (a) is for $^1 B$ state and (b) is for $^1 A$ one.}
\end{figure}

\begin{table}
\caption{Interchain parameters ($t_{\perp}$ and
$\tilde{U}$) and resulting exciton properties ($E_{\rm ex}$ and $P$) 
of
the lowest $^1 B$ states for the case of parallel ordering
between the two chains.}
\begin{tabular}{cccccccc}
&$t_{\perp},~\tilde{U}$ $(t)$ & 0.03, 0.5 & 0.03, 1.0 & 0.09, 0.5 & 
0.09,
1.0 & 0.15, 0.5 & 0.15, 1.0\\
\hline
&$E_{\rm ex}$ $(t)$ &1.095 &1.057  & 1.010 & 0.941 & 0.902 & 0.821\\
&$P$ &0.062 & 0.300 & 0.241 &0.428 & 0.330 & 0.457\\
\end{tabular}
\end{table}
\begin{table}
\caption{Interchain parameters ($t_{\perp}$ and
$\tilde{U}$) and resulting exciton properties ($E_{\rm ex}$ and $P$)
of the lowest $^1 A$ states for the case of parallel ordering
between the two chains.}
\begin{tabular}{cccccccc}
 &$t_{\perp},~\tilde{U}$ $(t)$ & 0.03, 0.5 & 0.03, 1.0 & 0.09, 0.5 &
 0.09, 1.0 & 0.15, 0.5 & 0.15, 1.0\\
\hline
 &$E_{\rm ex}$ $(t)$ &1.345 &1.331 & 1.229  &1.211 &1.109 & 1.091\\
 &$P$ &0.335 & 0.463 & 0.446 &0.488 & 0.468 & 0.493\\
\end{tabular}
\end{table}
\begin{table}
\caption{Interchain parameters ($t_{\perp}$ and
$\tilde{U}$) and resulting exciton properties ($E_{\rm ex}$,
$P$, and $d$) of the
lowest $^1 B$ states for the case of anti-parallel ordering.}
\begin{tabular}{cccccccc}
&$t_{\perp},~\tilde{U}$ $(t)$ & 0.03, 0.5 & 0.03, 1.0 & 0.09, 0.5 & 
0.09,
1.0 & 0.15, 0.5 & 0.15, 1.0\\
\hline
&$E_{\rm ex}$ $(t)$ &1.110 & 1.096 & 1.110 & 1.095 &1.108 & 1.093 \\
&$P$ & 0.005 & 0.007 & 0.045 & 0.059 & 0.111 & 0.139 \\
&$d$ $(a)$ &7 & 7 & 7 &7 & 7 & 7 \\
\end{tabular}
\end{table}
\begin{table}
\caption{Interchain parameters ($t_{\perp}$ and
$\tilde{U}$) and resulting exciton properties ($E_{\rm ex}$,
$P$, and $d$) of the
lowest $^1 A$ states for the case of anti-parallel ordering.}
\begin{tabular}{cccccccc}
&$t_{\perp},~\tilde{U}$ $(t)$ & 0.03, 0.5 & 0.03, 1.0 & 0.09, 0.5 & 
0.09,
1.0 & 0.15, 0.5 & 0.15, 1.0\\
\hline
&$E_{\rm ex}$ $(t)$ &1.345 & 1.194 & 1.328 & 1.188 & 1.308  & 1.178 \\
&$P$ &0.913 & 0.987 & 0.701 & 0.903 & 0.609 & 0.798\\
&$d$ $(a)$ & 9 & 7 & 9 & 7 & 7 & 5\\
\end{tabular}
\end{table}

\end{document}